\documentclass[a4paper,fleqn,usenatbib]{mnras}

\usepackage[T1]{fontenc}
\usepackage{ae,aecompl}

\usepackage{graphicx}	
\usepackage{amsmath}	
\usepackage{amssymb}	

\def\noao{The NOAO Deep Wide Field Survey MOSAIC Data Reductions, 
http://www.noao.edu/noao/noaodeep/ReductionOpt/frames.html}%
\title[LMC star clusters]{Star cluster formation history along the minor axis of the 
Large Magellanic Cloud}

\author[Piatti et al.]{
Andr\'es E. Piatti$^{1,2}$\thanks{E-mail: andres@oac.unc.edu.ar},
Andrew A. Cole$^3$ and Bryn Emptage$^3$
\\
$^{1}$Consejo Nacional de Investigaciones Cient\'{\i}ficas y T\'ecnicas, Av. Rivadavia 1917, 
C1033AAJ, Buenos Aires, Argentina\\
$^{2}$Observatorio Astron\'omico, Universidad Nacional de C\'ordoba, Laprida 854, 5000, 
C\'ordoba, Argentina\\
$^3$School of Physical Sciences, University of Tasmania, Private Bag 37, Hobart, 7001 TAS, Australia\\
}

\date{Accepted XXX. Received YYY; in original form ZZZ}

\pubyear{2017}

\begin{document}
\label{firstpage}
\pagerange{\pageref{firstpage}--\pageref{lastpage}}
\maketitle

\begin{abstract}
We analysed Washington $CMT_1$ photometry of star clusters located 
along the minor axis of the LMC, from the LMC optical centre
up to $\sim$ 39 degrees outwards to the North-West. The data base was exploited in order to search for new star cluster candidates,
to produce cluster CMDs cleaned from field star contamination and to derive age
estimates for a statistically complete cluster sample.
We confirmed  that 146 star
cluster candidates are genuine physical systems, 
and concluded that an overall $\sim$
30 per cent of catalogued clusters in the surveyed regions are unlikely to be true 
physical systems. 
We did not find any new cluster candidates in the outskirts of the LMC (deprojected distance
$\ga$ 8 degrees). The derived ages of the studied clusters are in the range 7.2
< log($t$ yr$^{-1}$) $\le$ 9.4, with the sole exception of the globular cluster 
NGC\,1786 (log($t$ yr$^{-1}$) = 10.10).
We also calculated the cluster frequency for each region, from which we confirmed previously 
proposed outside-in 
formation scenarios. In addition, we found that the outer LMC fields show a 
sudden episode of cluster formation (log($t$ yr$^{-1}$) $\sim$ 7.8-7.9) that continued until
log($t$ yr$^{-1}$) $\sim$ 7.3 only in the outermost LMC region. We link these features
to the first pericentre passage of the LMC to the MW, which could have triggered
cluster formation due to ram pressure interaction between the LMC and MW halo.
\end{abstract}

\begin{keywords}
techniques: photometric -- galaxies: individual: LMC --
galaxies: star clusters: general 
\end{keywords}



\section{Introduction}

Star clusters are invaluable probes of the structure and evolution of nearby galaxies. 
  Ever since the pioneering work of \citet{h60,h61}, it has been recognised that the 
  rich star clusters of the Large Magellanic Cloud (LMC) have the potential to reveal a 
  wealth of information about the star-formation, chemical evolution, and interaction history 
  of the largest nearby neighbour of the Milky Way. Rather early on it was already apparent
  that the history of the LMC, as traced by its clusters, is dramatically different from
  the history of the Milky Way \citep[e.g.,][]{swb80,fmb90}; the importance of a several 
  billion year-long hiatus in cluster formation was emphasised by \citet{dc91}.  
  
 Due to the proximity of the LMC to the Milky Way, and of the Small Magellanic Cloud (SMC) 
 to the LMC, the known LMC cluster age gap  \citep{richetal2001,bekkietal2004} and the 
 spatial and kinematic distributions of  LMC clusters on 
 its older and younger sides  \citep{scho1991,ee1998} have been the subject of much attention in studies of 
 the tidal
 and hydrodynamical interactions between the three galaxies \citep[e.g.,][]{fmb90,bch05}. 
 Because of evidence of tidal interactions between the Magellanic Clouds 
 \citep{beslaetal2016},
  the LMC could have experienced a much more complicated star-formation history than would be 
  inferred simply by assuming a constant ratio of stars in clusters to those in the field 
  \citep[e.g.,][]{getal98}, and that significant spatial variations exist, particular when 
  comparing the central bar to the disc \citep{shetal02,hz09}.

\newpage

 In recent years, numerous deep colour-magnitude diagrams of the LMC field have been
 published, giving a dramatically improved view of its long-term star-formation history
 \citep[e.g.,][and references therein]{wetal13}. Concurrently, there has been a revolution
 in our view of the orbital motion and interaction history of both Magellanic Clouds, 
 driven by new measurements of their proper motions \citep{kallivayaliletal13}. This new precision 
 promises the ability to reliably trace the correlations between tidal and hydrodynamical
 interactions, global and local enhancements in star formation rate, and the production and
 destruction of star clusters. 
 
 In this context, it is crucial to have a highly reliable catalogue of clusters, with 
 accurate and precise age and metallicity information. 
 Because the luminosity function of clusters increases steeply toward the faint end, it 
 is inevitable that there will be many more poorly-measured clusters than well-measured 
 ones, and every opportunity to revisit cluster parameters is potentially valuable
 \citep[e.g.][]{pc17,p17b}. In this paper, we examine the LMC cluster 
 population using deep imaging along the bar and minor axis of the disc. The original 
 data set was obtained to search for extremely metal-poor field stars (Emptage et al., 
 in preparation), but is also well-suited to cluster studies. Here, we 1) statistically 
 test the existence of suggested very low-mass clusters; 2) search for previously 
 unknown clusters; 3) refine the measured ages and metallicities of understudied 
 clusters; and 4) examine the spatial-temporal properties of the cluster population in 
 this long and narrow strip of sky.

The paper is organized as follows: in Section 2 we describe the data processing
and the standardization of the obtained stellar photometry. Section 3 deals with
the compilation of a statistically complete cluster sample from this data set,
which comprises the search for new star clusters and the cleaning of the cluster
CMDs from field star contamination. In Sections 4 and 5 
we estimate the cluster ages and studied the cluster formation history along the
minor axis of the LMC, respectively. Finally, Section 5 summarizes the main 
conclusions of our work.

\section{Data processing and standardization}

\begin{figure*}
\includegraphics[width=\columnwidth]{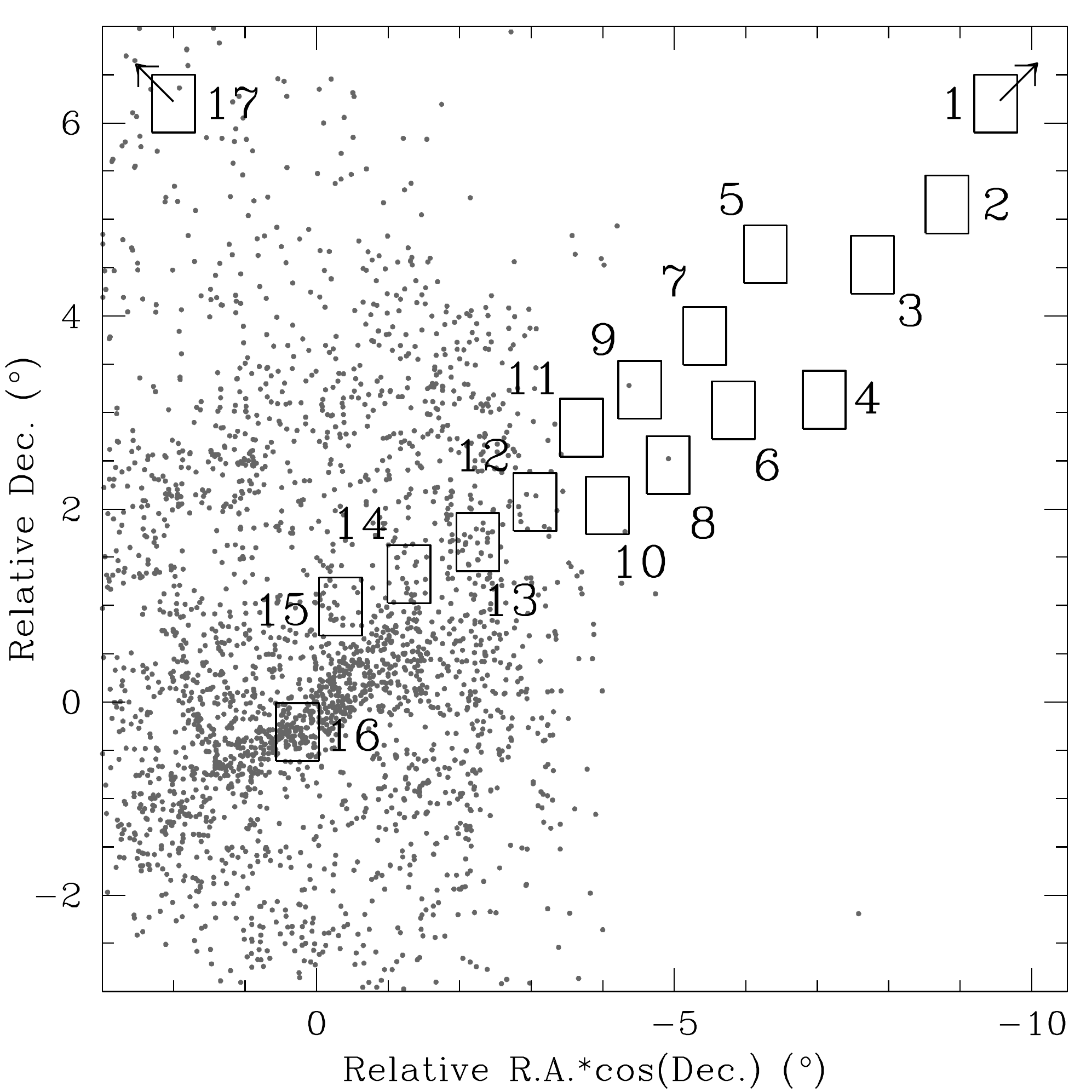}
\includegraphics[width=\columnwidth]{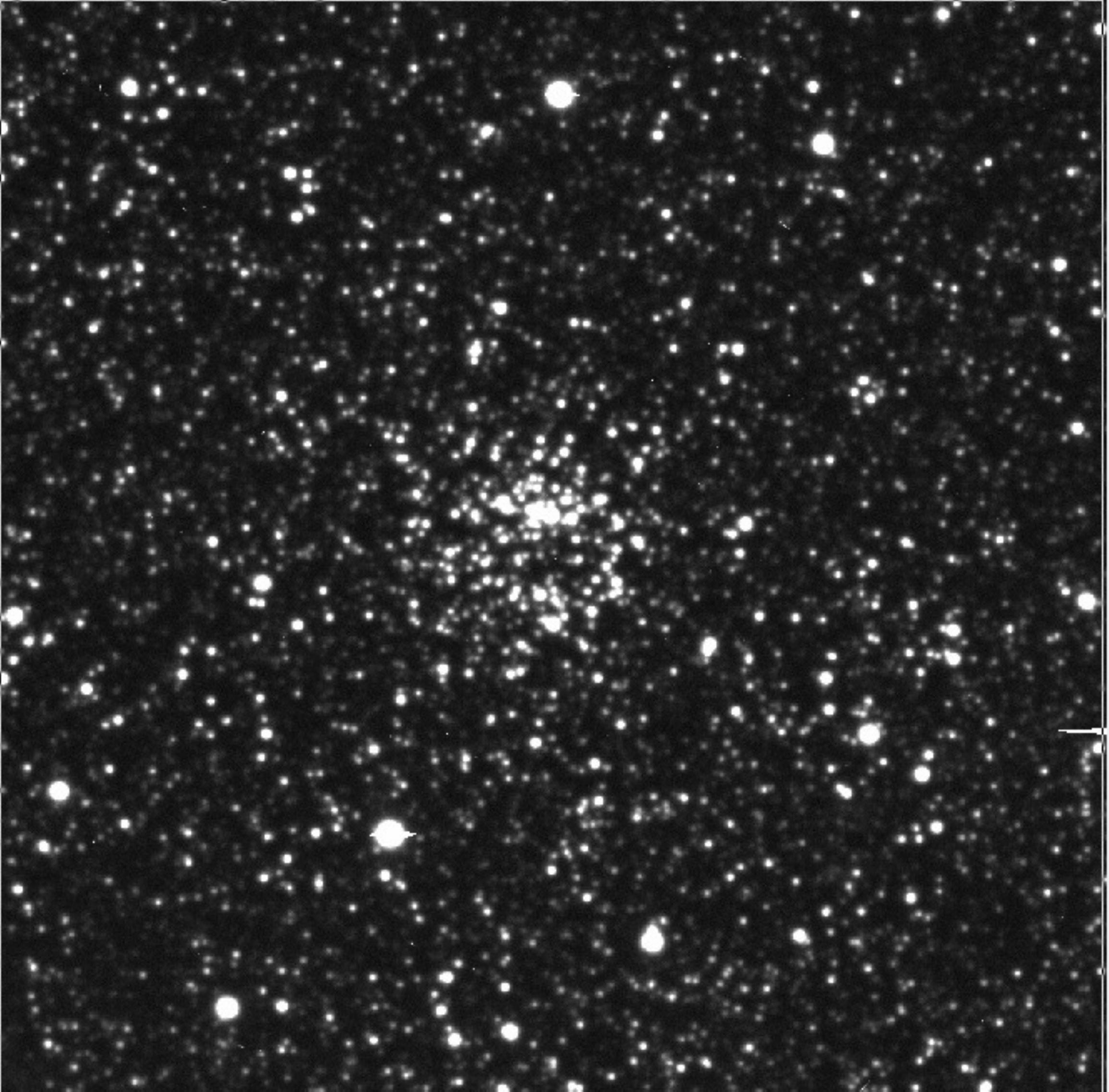}
\includegraphics[width=\columnwidth]{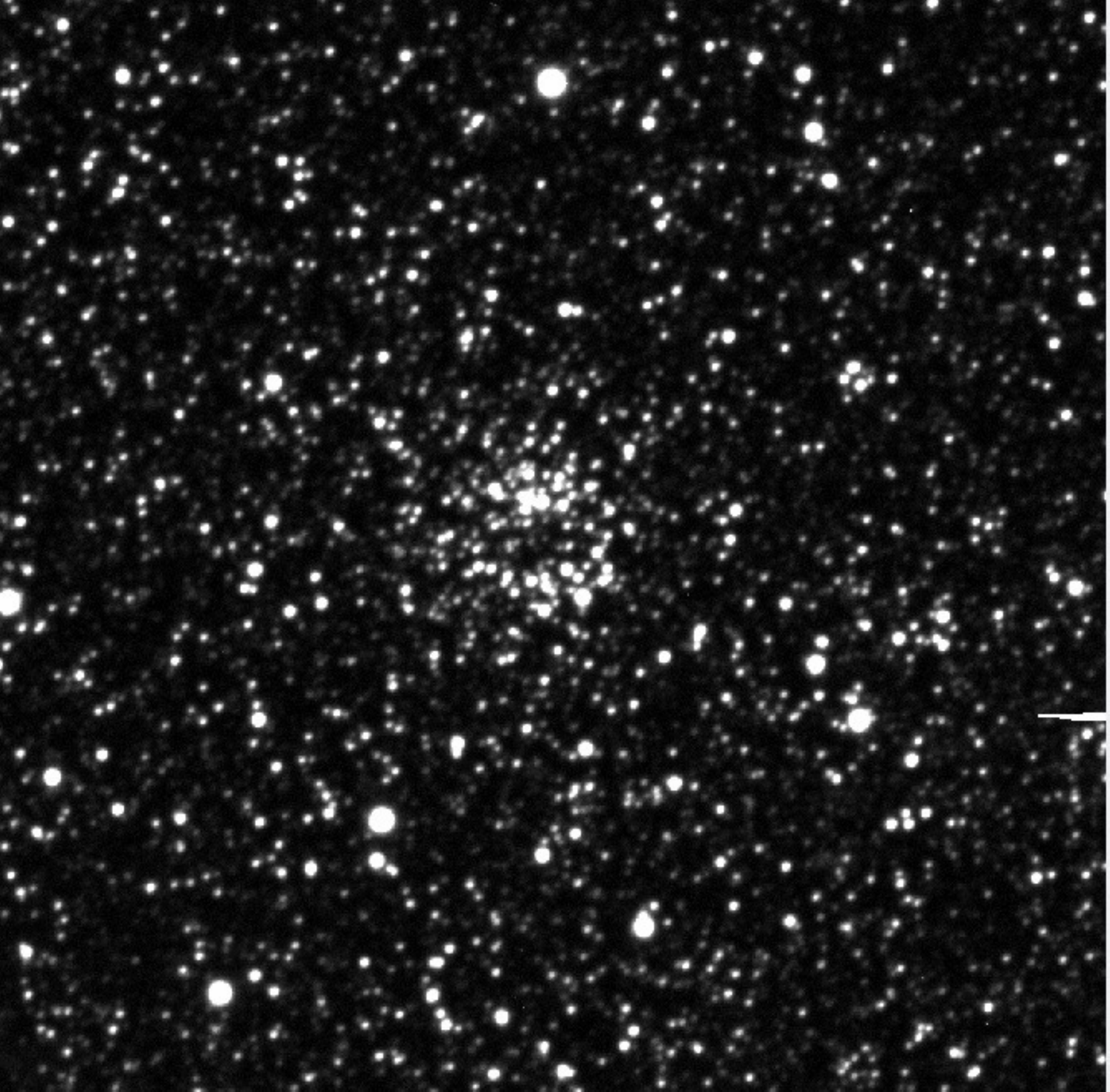}
\includegraphics[width=\columnwidth]{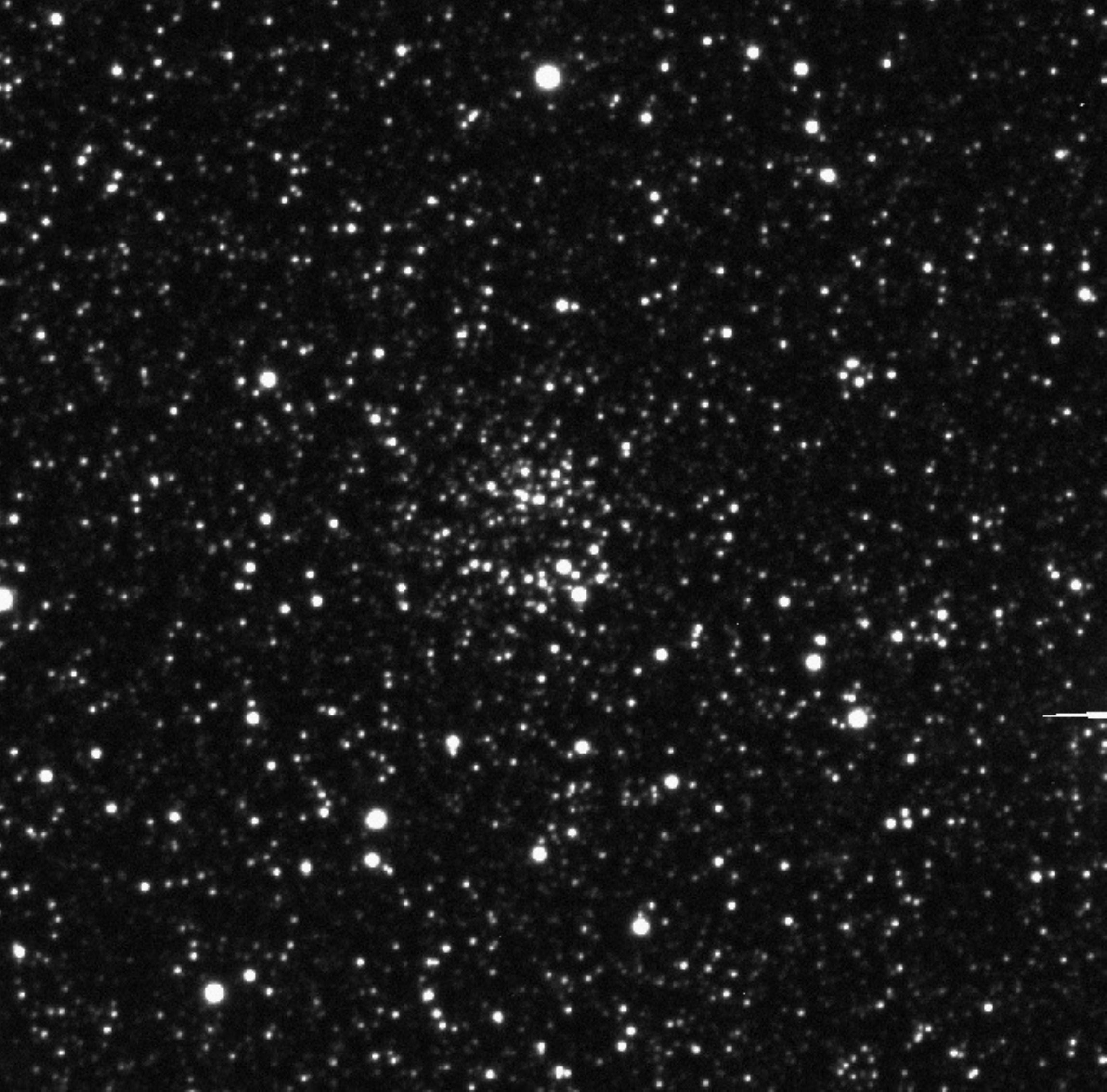}
    \caption{Top-left: Spatial distribution of the presently studied LMC star fields (thick black
    boxes). Star clusters catalogued by \citet{betal08} are also drawn (dots) for
comparison purposes. To illustrate the data quality we include 4$\times$4 arcmin subsections of our $CMR$ images of Field\,15, centred on field BRHT\,34a,b 
in the bottom-left, top-right and 
bottom-right panels, respectively. North is up and East to the left. The full $CMR$ images of
Field\,15 (the most densely populated one, regardless Field\,16 \citep{p17b}), and those for Field\,1
(the least populated one) are provided as supplementary material in the online version of the
journal.}
   \label{fig:fig1}
\end{figure*}

The data used in this work come from the Cerro-Tololo Inter-American Observatory 
(CTIO) programme 2008B-0296 (PI: Cole) focused on surveying the most metal-poor 
stars outside the Milky Way. We downloaded from the National Optical Astronomy 
Observatory  (NOAO) Science Data Management (SDM) 
Archives\footnote{http://www.noao.edu/sdm/archives.php.} Washington $CM$ and 
Kron-Cousins $R$ images
obtained with the Mosaic\,II imager, an array of 8K$\times$8K CCDs
covering a 36$\arcmin$$\times$36$\arcmin$ field, attached to the 4 m Blanco telescope.
Table~\ref{tab:table1} presents the log of the observations, where the main 
astrometric and observational information is summarized. The whole survey comprises 
17 different LMC fields as illustrated in Fig.~\ref{fig:fig1}.


As previously performed for similar data sets 
\citep[e.g.][and references therein]{pietal12,p12a,p15}, we carried out the data processing 
and obtained the standardized
photometry  following the procedures documented by the NOAO Deep Wide 
Field Survey team \citep{jetal03}, and utilizing the {\sc mscred} package in 
IRAF\footnote{IRAF is distributed by the National 
Optical Astronomy Observatories, which is operated by the Association of 
Universities for Research in Astronomy, Inc., under contract with the National 
Science Foundation.} and the {\sc daophot/allstar} suite of programs \citep{setal90}.
The raw images were first customized by performing overscan, trimming, bias subtraction 
and flat-field corrections. The applied zero and sky- and 
dome- flats come from properly combined individual ones. We also took advantage
of  $\sim$ 500 stars catalogued by the 
USNO\footnote{http://www.usno.navy.mil/USNO/astrometry/optical-IR-prod/icas/usno-icas}
to obtain an updated world coordinate system (WCS) with an rms error smaller than 0.4 
arcsec in RA and DEC.

We measured nearly 270 independent 
magnitudes in the standard fields PG0321+051, SA\,98 and SA\,101 \citep{l92,g96} -- observed
three times per night  (Dec. 27 -- 30, 2008) -- using the {\sc apphot} task within 
IRAF. These magnitudes were used to derive the coefficients to transform 
the instrumental $cmr$ system to the Washington $CMT_1$ system. We fitted the
following expressions:

\begin{equation}
c = c_1 + T_1 + (C-T_1) + c_2\times X_C + c_3\times (C-T_1),
\end{equation}

\begin{equation}
m = m_1 + T_1 + (M-T_1) + m_2\times X_M + m_3\times (M-T_1),
\end{equation}

\begin{equation}
r = {t_1}_1 + T_1 + {t_1}_2\times X_{T_1} + {t_1}_3\times (C-T_1),
\end{equation}

\noindent where  $c_i$, $m_i$ and ${t_1}_i$ ($i$ = 1, 2 and 3) are the fitted 
coefficients, and $X$ represents the effective airmass. Instrumental and
standard magnitudes are distinguished by lowercase and capital letters,
respectively. Note that eq. (3) involves $r$ magnitudes to derive $T_1$ magnitudes 
because the $R$ filter is the recommended  substitute of the Washington
$T_1$ filter \citep{g96}. The resultant transformation
coefficients for each night, obtained with the {\sc fitparams} task in IRAF,
are shown in Table~\ref{tab:table2}.

The stellar photometry for each single mosaic --
produced by gathering all together the 8 CCDs using the updated WCS -- was obtained
after deriving the respective quadratically varying point-spread-function (PSF). Such a PSF was created
by using two lists of stars, one with $\sim$ 1000 and another with the brightest 
$\sim$ 250 stars, both selected interactively. A preliminary PSF is obtained from 
the smallest sample, which is used to clean the largest PSF star sample.
The procedure to derive PSF magnitudes of the stars identified in each field
consisted in applying the resultant PSF to the single mosaic; then identifying
new fainter stars in the output subtracted frame, and running again the {\sc allstar} 
program for the enlarged sample of stars. We iterated this loop three times.
Only objects with $\chi$ $<$ 2, photometric error less than 2$\sigma$ above the 
mean error at a given magnitude, roundness values between 
-0.5 and 0.5 and sharpness values between 0.2 and 1.0 were kept.
Finally, we used eqs. (1) to (3) to standardize the PSF instrumental magnitudes
and the {\sc daomatch} 
and {\sc daomaster} programs\footnote{Provided kindly by Peter Stetson.} to 
put the stellar $CMT_1$ magnitudes of each field into a single file.

\begin{figure*}
	\includegraphics[width=\textwidth]{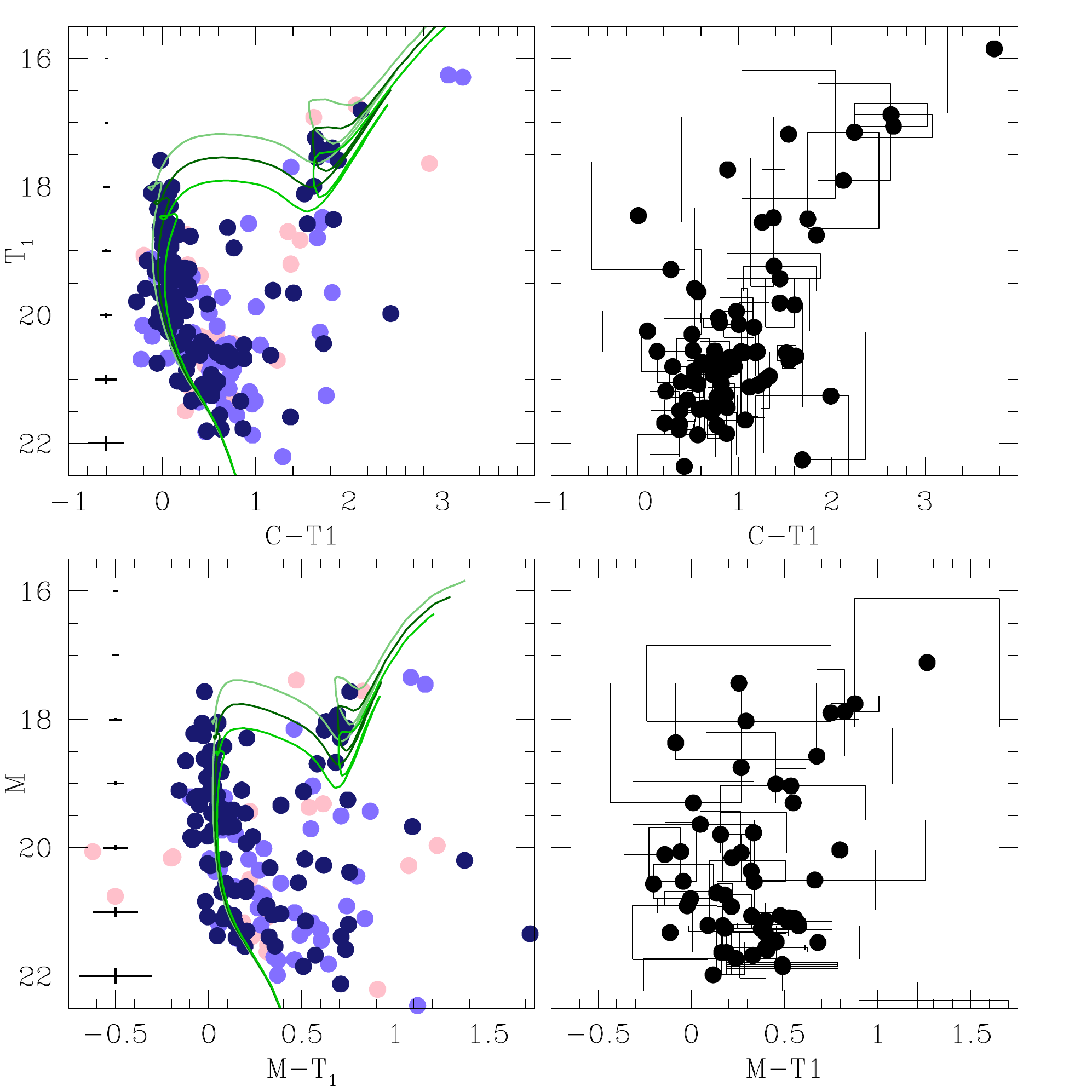}
    \caption{CMDs for stars in the field of KMHK\,691  with typical
 photometric errors represented with errorbars at the left margin
 Colour-scaled symbols represent stars that statistically
  belong to the field ($P \le$ 25\%, pink), stars that might belong to
  either the field or the cluster ($P =$ 50\%, light blue), and stars
  that predominantly populate the cluster region ($P \ge$ 75\%, dark
  blue). 
  Three isochrones from \citet{betal12} for log($t$ yr$^{-1}$)
  = 8.5 (pale green), 8.6 (dark green), and 8.7 (medium green) and $Z$ = 0.006 are also superimposed.  Right panels
  illustrate the definition of boxes in the field-star CMD.}
   \label{fig:fig2}
\end{figure*}

We estimated the errors  of our photometry from artificial
star tests carried out using the stand-alone {\sc addstar} program in the {\sc daophot}
package \cite{setal90} to add synthetic stars with Poisson noise, 
generated bearing in mind the colour and magnitude distributions 
of the stars in the cluster colour-magnitude diagrams (CMDs), as well as their radial 
stellar density profiles. We added  $\sim$ 5$\%$ of the measured stars in order
to produce a thousand synthetic images with similar stellar densities as observed. The synthetic images were used to obtain stellar 
PSF magnitudes as described above. Then,  by comparing the output 
and the input magnitudes of the added stars we estimated the respective
photometric errors. Fig.~\ref{fig:fig2} illustrates the typical photometric errors 
with errorbars at the left margin the CMDs.

\begin{table*}
\caption{Observations log of selected LMC star fields.}
\label{tab:table1}
\begin{tabular}{@{}lccccccccccccccc}\hline
ID  &R.A.(J2000.0)      &Dec.(J2000.0)     & \multicolumn{3}{c}{filters$^a$} & \multicolumn{3}{c}{exposures} & \multicolumn{3}{c}{airmass} & \multicolumn{3}{c}{mean seeing} \\
         &(h m s)   &($\degr$ $\arcmin$ $\arcsec$)&  \multicolumn{3}{c}{} &  \multicolumn{3}{c}{(sec)} &  \multicolumn{3}{c}{}& \multicolumn{3}{c}{($\arcsec$)} \\
\hline

Field\,1  & 01 09 34.28 & -52 23 10.3 & C& M& R & 1$\times$150& 1$\times$20& 1$\times$15 & 1.22& 1.22 &1.23 & 1.2& 1.1& 1.1 \\
Field\,2  & 03 59 33.04 & -64 19 29.3 & C& M& R & 1$\times$300 &1$\times$60 &1$\times$30 & 1.22& 1-22& 1.23 & 1.3& 1.2& 1.2 \\
Field\,3  & 04 07 30.02 & -64 56 48.8 & C& M& R & 1$\times$420& 1$\times$60 &1$\times$30 & 1.22& 1.22& 1.23 & 1.0 &1.0& 1.0 \\
Field\,4  & 04 10 09.88 & -66 20 58.6 & C& M& R & 1$\times$300 &1$\times$60& 1$\times$30 & 1.24& 1.24& 1.24 & 1.1&1.0 &1.0 \\
          &             &             &  C& M& R & 1$\times$300& 1$\times$45 &1$\times$30 & 1.30 &1.30& 1.31 & 1.2& 1.2& 1.1 \\
Field\,5  & 04 21 53.04 & -64 50 26.2 & C& M& R & 1$\times$300& 1$\times$30&1$\times$20 & 1.29& 1.30 &1.29 & 1.3 &1.2 &1.2 \\
Field\,6  & 04 22 35.71 & -66 27 25.9 & --&M& R   & --&  1$\times$120& 1$\times$30         & -- &  1.24 &1.24 & --& 1.0& 1.0 \\
Field\,7  & 04 28 13.14 & -65 41 13.6 & --&M& R   & --  & 1$\times$60& 1$\times$30         & -- &  1.24& 1.24 & --& 1.1& 1.1 \\
Field\,8  & 04 30 36.03 & -67 01 26.0 & C& --&R   & 1$\times$420& --& 1$\times$30          & 1.25 &--  & 1.25 & 1.2& --& 1.1 \\
Field\,9  & 04 36 03.74 & -66 14 30.8 & --&M& R   & -- &1$\times$60& 1$\times$30           & -- &  1.34& 1.34 & --& 1.0& 1.0 \\
Field\,10 & 04 38 30.77 & -67 26 30.8 & --&M& R   & -- &1$\times$45& 1$\times$20           & -- &  1.55&1.51 & --& 1.2& 1.2 \\
Field\,11 & 04 43 36.55 & -66 38 12.1 & --&M& R   & -- &1$\times$60& 1$\times$30           & -- &  1.27& 1.28 & -- &1.2& 1.2 \\
Field\,12 & 04 49 11.12 & -67 24 22.0 & --&M& R   & -- & 1$\times$60& 1$\times$30          & -- &  1.29& 1.29 & --& 1.1 &1.1 \\
Field\,13 & 04 57 04.68 & -67 49 17.4 & --&M& R   & -- & 1$\times$60 &1$\times$30          & -- &  1.63& 1.64 & -- &1.1& 1.1 \\
Field\,14 & 05 07 03.50 & -68 09 19.1 & --&M& R   & --& 1$\times$60& 1$\times$30           & --  & 1.45&1.44 & -- &1.1& 1.0 \\
Field\,15 & 05 17 18.89 & -68 29 14.6 & C& M& R & 1$\times$420& 1$\times$60& 1$\times$30 & 1.58 &1.57& 1.61 & 1.3 &1.2 &1.2 \\
Field\,16 & 05 24 04.21 & -69 47 23.8 & C& M& R & 1$\times$420& 1$\times$60 &1$\times$30 & 1.34 &1.34& 1.31 & 1.2& 1.2 &1.2 \\
Field\,17 & 05 58 31.97 & -48 38 32.5 & C& M& R &  1$\times$80& 1$\times$20& 1$\times$10 & 1.39 &1.38 &1.41 & 1.1 &1.1&1.1 \\
\hline
\end{tabular}

\noindent $^a$ Note that the Kron-Counsins $R$ filter is the recommended substitute of the Washington
$T_1$ filter \citep{g96}.

\end{table*}

\begin{table}
\caption{Washington $CMT_1$ transformation ero point (1), extinction
(2) and colour term (3) coefficients.}
\label{tab:table2}
\begin{tabular}{@{}lcccc}\hline
Date (UT) & $c_1$       &     $c_2$      &    $c_3$      & rms  \\\hline
Dec. 27 &  0.149$\pm$0.013&   0.280$\pm$0.010&  -0.099$\pm$0.007&   0.037 \\
Dec. 28 &  0.033$\pm$0.017&   0.292$\pm$0.011&  -0.083$\pm$0.004&   0.030 \\
Dec. 29 &  0.030$\pm$0.019&   0.283$\pm$0.130&  -0.086$\pm$0.005&   0.029 \\
Dec. 30 &  0.016$\pm$0.015&   0.289$\pm$0.009&  -0.094$\pm$0.003&   0.023 \\\hline
    Date (UT)    & $m_1$       &     $m_2$      &    $m_3$      & rms \\\hline
Dec. 27 & -0.926$\pm$0.009 & 0.140$\pm$0.010 & -0.246$\pm$0.012 &  0.016 \\
Dec. 28 & -0.999$\pm$0.014 & 0.132$\pm$0.009 & -0.230$\pm$0.008 &  0.024\\
Dec. 29 & -1.028$\pm$0.015  &0.145$\pm$0.009 & -0.240$\pm$0.008 &  0.022\\
Dec. 30 & -1.034$\pm$0.013 & 0.139$\pm$0.008 & -0.234$\pm$0.007 &  0.020\\\hline
   Date (UT)     & ${t_1}_1$ & ${t_1}_2$ &  ${t_1}_3$ & rms \\\hline
Dec. 27 &-0.605$\pm$0.010 & 0.080$\pm$0.010 & -0.030$\pm$0.006  &  0.031\\
Dec. 28 &-0.679$\pm$0.015 & 0.080$\pm$0.011 & -0.018$\pm$0.003  &  0.021\\
Dec. 29 & -0.665$\pm$0.019 & 0.071$\pm$0.013 & -0.026$\pm$0.004 &   0.029\\
Dec. 30 & -0.723$\pm$0.008 & 0.101$\pm$0.005 & -0.025$\pm$0.002  &  0.013\\
\hline
\end{tabular}
\end{table}

\section{The star cluster sample}

Instead of using the up-to-date list of catalogued clusters, we decided
to perform an homogeneous search over all observed LMC fields using 
the procedure developed in \citet{petal2016}
and also successfully used elsewhere \citep[e.g.][]{p16,p17}.
Thus, we could not only recover the known clusters but also search
for new ones, particularly extending to the fields beyond the LMC main body.
For the sake of the reader we describe the steps followed: we started 
by using the {\sc daofind} task within
{\sc daophot} to detect every stellar source in the deepest images.
Then, we built continuous density distributions using 
two different Kernel Density Estimators (KDEs),  namely, {\it Gaussian} 
and {\it tophat}, and a KDE bandwidth of 0.4 arcmin, that allowed
us to extract the finest structures of them (e.g., smallest
and/or less dense resolved clusters). \citet{petal2016} showed that
the mean stellar cluster density as a function of cluster radius is a
suitable diagnostic diagram to infer the appropriate bandwidth;
the only free parameter while using KDE. Such a diagram shows the range of cluster
sizes and their stellar densities, so that in order to detect the smallest
clusters, a KDE bandwidth of the order of the diameter of the smallest 
clusters should be used. They showed that such a criterion allowed them
to recover 100 per cent of the known clusters \citep[see also,][]{p17,p17b}. 
By using larger bandwidths, some small clusters could be missed.

To this purpose, we used
the Python KDE routing within AstroML 
\citep[][and reference therein for a detail
description of the complete AstroML package and user's
manual]{astroml}, a machine learning and data mining for Astronomy
package. AstroML is a Python module for machine learning and data mining 
built on {\it numpy, scipy, scikit-learn, matplotlib, and {\it astropy},
and distributed under the 3-clause BSD license. It contains a growing 
library of statistical and machine learning routines for analysing 
astronomical data in Python, loaders for several open astronomical 
data sets, and a large suite of examples of analysing and visualizing
astronomical datasets.
The goal of AstroML is to provide a community repository for fast Python 
implementations of common tools and routines used for statistical data
analysis in astronomy and astrophysics, to provide a uniform and
easy-to-use interface to freely available astronomical data sets.}

The second step consisted in deciding which of the total number of overdensities 
detected above are star cluster candidates. Here, we considered the height 
(i.e., the peak in stellar density of an overdensity) relative
to the local background and a cut-off density of 
1.5 times the local background dispersion above the mean
background value, to increase the chance of identifying 
star cluster candidates \citep{petal2016}. As far as we are aware,
we used the most suitable values in order
to identify star cluster candidates, since we chose a bandwidth value of the order
of the diameter of the smallest known clusters. Indeed, we identified all the
previously known star clusters and detected some new ones. By using smallest
bandwidths, the same number of true clusters is recovered. The appropriate
height relative to the local background was adopted by looking at the stellar
density versus local background plot for a grid of points distributed throughout the 
field.
\citet{petal2016} assumed similar conditions
while searching for new star clusters in the densest and
highest reddened region ($\sim$ 0.4 deg$^2$) of the Small Magellanic Cloud (SMC) bar, 
and identified the 68 catalogued star clusters in \citet[][hereafter B08]{betal08} and 38 new
ones. Our resulting numbers of star cluster candidates are listed in
Table~\ref{tab:table3}.

In order to decontaminate the star cluster candidate CMDs from field stars we
applied a procedure developed by \citet{pb12}, and successfully used elsewhere 
\citep[e.g.][and references therein]{p14,petal15a,petal15b,pb16a}.
Briefly, the field star cleaning relies on the subtraction of different
 previously defined field-star CMDs to the cluster CMD. We used a total of four
 different field-star CMDs of regions located to the north, east, south and west from
 the cluster centre, respectively. The field regions have the same area as that used 
 for the cluster, i.e., of a circle of radius three times as big as the cluster radii 
 given by B08. For each field-star CMD we produced a number of boxes -- as many as
stars in  the field-star CMD -- that are centred at
the magnitudes and colours of the  stars and that have sizes as large
as their corners coincide with the position of their closest star; magnitude and colour
sides are adjusted independently.  Right panels of
Fig.~\ref{fig:fig2} illustrate the definition of the boxes in the field-star CMD.
We then placed those boxes on the cluster CMD and
eliminated one star per box, choosing the closest one to its centre that lies inside it.
We repeated the procedure for all the four field-star CMDs.
The method has proved to be powerful to effectively reproduce the  local field star
signature in terms of stellar density, luminosity function and colour distribution.
Here we used four field star CMDs constructed from stars 
located around the clusters and with areas equal to the circular area used for
the cluster region.

Photometric membership probabilities were then assigned on the basis of the number
of times a star in the cluster CMD  was kept unsubtracted, once the four independent 
 cleanings
were executed. If a star appeared once in the four cleaned CMDs (the same cluster CMD cleaned
from four different field-star CMDs), we gave it a membership
probability  $P \le$ 25 per cent. If a star appeared twice, then it was given a probability
$P =$ 50 per cent; and for more than twice, $P \ge$ 75 per cent.
Because of the statistical subtraction process,
a small amount of residuals can be arise, 
depending  on the particular features of the 
considered star field (stellar density variability, etc). 
We finally discarded two objects known in the literature as [HS66]\,197 and 
KMHK309n, whose cleaned CMDs do not show any detectable trace of star cluster sequences.
Table~\ref{tab:table3} presents the statistics of confirmed star cluster per field,
while Figure~\ref{fig:fig2} illustrates the performance of the cleaning
procedure for KMHK\,691.
The individual photometric catalogues for the confirmed clusters are
provided in the online version of the journal. The columns of each
catalogue successively lists the star ID, the R.A. and Dec., the X and Y coordinates
(in pixels), the magnitude
and error in $C$, $M$ and $T_1$, respectively, $\chi$ and sharpness, and the photometric 
membership probability ($P$). The latter is encoded with numbers 1, 2, 3 and 4 to
represent probabilities of 25, 50, 75 and 100 per cent, respectively.
 A portion of the photometric catalogue of KMHK\,685
is shown in the Appendix
for guidance regarding their form and content.

As far as we are aware, the final cluster sample is statistically complete, 
if we bear in mind the depth of our photometry and the distance of the LMC.
Indeed, young star clusters are distinguished in the CMDs by their bright MSs, 
while intermediate-age clusters have
main-sequence turnoffs (MSTOs) that decrease in brightness as they become older.
The oldest intermediate-age clusters are $\sim$ 2.5 Gyr old \citep{pg13}, which in turn implies
a MSTO at $T_1$ $\sim$ (20.5 $\pm$ 0.15) mag, assuming an average depth of
3.44$\pm$1.16 kpc \citep{ss09}. 
This magnitude is  brighter than our limiting magnitude, so that we were able to detect any star cluster based on counts  of its brightest stars all the way down to its MSTO. The exception to this would be genuinely old globular clusters 
 (log($t$ yr$^{-1}$) $\sim$ 10), which have fainter MSTOs.

Nevertheless, we performed a matching between the catalogued clusters
by B08 and \citet{getal10} and our final cluster  sample. 
We confirmed that every object
in our sample is in B08/\citet{getal10}, except two new cluster candidates, one 
reported by \citet{p17b} and another one with coordinates 
R.A.= 72$\degr$.579697, DEC.= -67$\degr$.703239  (J2000.0). The latter 
was not resolved by the SIMBAD\footnote{http://simbad.u-strasbg.fr/simbad/}
astronomical data base either. The very few new star cluster candidates
detected would not appear to support some recent outcomes that have shown 
that there are still a substantial number 
of extreme low luminosity stellar clusters undetected in the wider Magellanic
System periphery and the
Milky Way halo \citep{kimetal2015a,pieresetal2016,martinetal2016}. 
These latter results are based on deep images obtained with the Dark Energy Camera 
(DECam) at the CTIO 4 m
Blanco telescope \citep[see][]{valdesetal2014}, and the discovery of some few faint 
extended stellar objects led to speculate on the possible existence of
larger amounts of star clusters.
Furthermore, streams
of gas and stars that might harbour stellar clusters have also been detected
\citep{mackeyetal2016,belokurovetal2017,deasonetal2017}, but a complete search
for new clusters in the DECam fields is still pending.  Conversely, our
findings agrees well with other recent results obtained by \citet{p17}, who
also concluded on low chances of detecting a significant number of stellar clusters
there  from DECam deep images.
 
We also found that several B08's clusters  were not identified by our procedure; 
one of them (H88\,34) because it falls on a Mosaic II image gap. 
The remaining objects could 
not be  recognised when visually inspecting the $C$, $M$ and $T_1$ images  either, because 
the distribution of stars in their respective fields do not resemble that of 
a stellar aggregate. We consider them as probable  non-genuine star clusters. 
They are: KMHK\,125 in Field\,12; BSDL\,616, 631, 661, 677, 
[GKK2003]\,O219, O222, KMHK\, 609 and OGLE-CL\,LMC\,122 in Field\,14; BSDL\, 962, 1157, 
1294, [HS66]\,231 and KMHK759 in Field\,15. For Field\,16, we refer the reader to 
\citet{p17b}, who also discusses the origin of those asterisms, namely, the
lower spatial resolution and magnitude limit \citep[see also,][]{pb12,p14}.

\begin{table}
\caption{Statistics of star clusters in the presently studied LMC star fields. }
\label{tab:table3}
\begin{tabular}{@{}lccc}\hline
ID & detected star  &  confirmed   &  B08 unlikely \\
   & cluster candidates & star clusters& star clusters \\\hline

Field\,1  & -- & -- & -- \\
Field\,2  & -- & -- & -- \\
Field\,3  & -- & -- & -- \\
Field\,4  & -- & -- & -- \\
Field\,5  & -- & -- & -- \\
Field\,6  & -- & -- & -- \\
Field\,7  & -- & -- & -- \\
Field\,8  & 1 & 1 & -- \\
Field\,9  & 1 & 1 & -- \\
Field\,10 & 1 & 1 & -- \\
Field\,11 & 1 & 1 & -- \\
Field\,12 & 11 & 11  & 1 \\
Field\,13 & 24$^a$ & 23 & --\\
Field\,14 & 17 & 17  & 8 \\
Field\,15 & 22 & 21 & 5 \\
Field\,16$^b$ & 73 & 70 & 38 \\ 
Field\,17 & -- & -- & -- \\
\hline
\end{tabular}

\noindent $^a$ H88\,34 was not detected, because it falls on an image gap.

\noindent $^b$ values taken from \citet{p17b}.
\end{table}

\section{Star cluster age estimates}

In order to estimate the ages of the studied clusters, we used the theoretical 
isochrones of \citet{betal12} to match their
CMDs built from stars with membership probabilities higher than 50 per cent. 
Hence we assigned
to a cluster an age equal to the isochrone's age which best resembles the
cluster CMD features. Because of the LMC distance 
\citep[49.90$^{-2.04}_{+2.10}$ kpc, ][]{dgetal14} and its depth \citep[3.44$\pm$1.16 kpc, ][]{ss09},
the difference in the cluster distance moduli could be as large as $\Delta$$(m-M)_o$ $\sim$ 
0.3 mag. This difference is similar to that we would get at the MSTO $T_1$ mag 
when matching theoretical isochrones to the cluster CMDs, aiming at reproducing the
observed dispersion (see Fig.~\ref{fig:fig2}). The inclination of the LMC disc to the line
of sight could also contribute to distance differences, but the position angle of the observed
fields is near to the line of nodes \citep[e.g.][]{vdmarel2001}, so the distance change
is unlikely to be large compared to its uncertainty. For this reason, we adopted a mean
distance modulus of $(m-M)_o$ = 18.49 mag for all the clusters. 

We also used isochrones for Z = 0.006 ([Fe/H] = $-$0.4 dex), which 
corresponds to the mean LMC metal content during the last $\sim$ 2-3 Gyr.
Note that  the age-metallicity relationship for LMC clusters derived by \citet{pg13} 
shows that the LMC chemical evolution has mostly taken place
within a constrained metallicity range during this period ([Fe/H] $\approx$ 
-0.7 dex to -0.2 dex) and differences in theoretical isochrones, particularly
along the main sequence (MS), are negligible compared to the observed dispersion.
We made one exception in the employment of the isochrones for the old globular 
cluster NGC\,1786,  that lies in Field\,13, and for which we adopted [Fe/H] = $-$2 dex 
\citep{brocatoetal1996,carrettaetal2000}.

Since the reddening is expected to vary across the surveyed regions, we
estimated $E(V-I)$ colour excesses from the Magellanic Clouds (MCs) extinction 
values based on the red clump (RC) and  RR Lyrae stellar photometry provided by 
the  Optical Gravitational Lens Experiment \citep[][ OGLE\,III]{u03} collaboration, 
as described in \citet{hetal11}. In matching the isochrones, we started by adopting 
those $E(V-I)$ values, combined with the equations 
$E(V-I)$/$E(B-V)$ = 1.25, $A_{V}$/$E(B-V)$ = 3.1 \citep{cetal89}, 
$E(C-T_1)$/$E(B-V)$ = 1.97 and $A_{T_1}$/$E(B-V)$ = 2.62 \citep{g96}, and
the adopted mean LMC distance modulus, to properly shift the theoretical isochrones
in $T_1$ and $M$ magnitudes and $C-T_1$ and $M-T_1$ colours.
When overplotting the theoretical isochrones, we used the shape of the MS, its curvature, 
the relative distance between the RC and the MSTO in magnitude and colour, separately,
among others, as reddening- and distance-free features to choose the isochrone which
best reproduce them. We estimated the overall age uncertainty associated to the observed 
dispersion  in magnitude in the cluster CMDs to be $\Delta$log($t$ yr$^{-1}$) = $\pm$0.10.
 Notice that the magnitude of the MSTO is age-dependent and that the position of
the red clump also constrains the age range.
Fig.~\ref{fig:fig2}  illustrates the performance of the isochrone matching for two
different CMDs, while Table~\ref{tab:table4} lists the derived $E(V-I)$
colour excesses and ages.

\begin{figure}
	\includegraphics[width=\columnwidth]{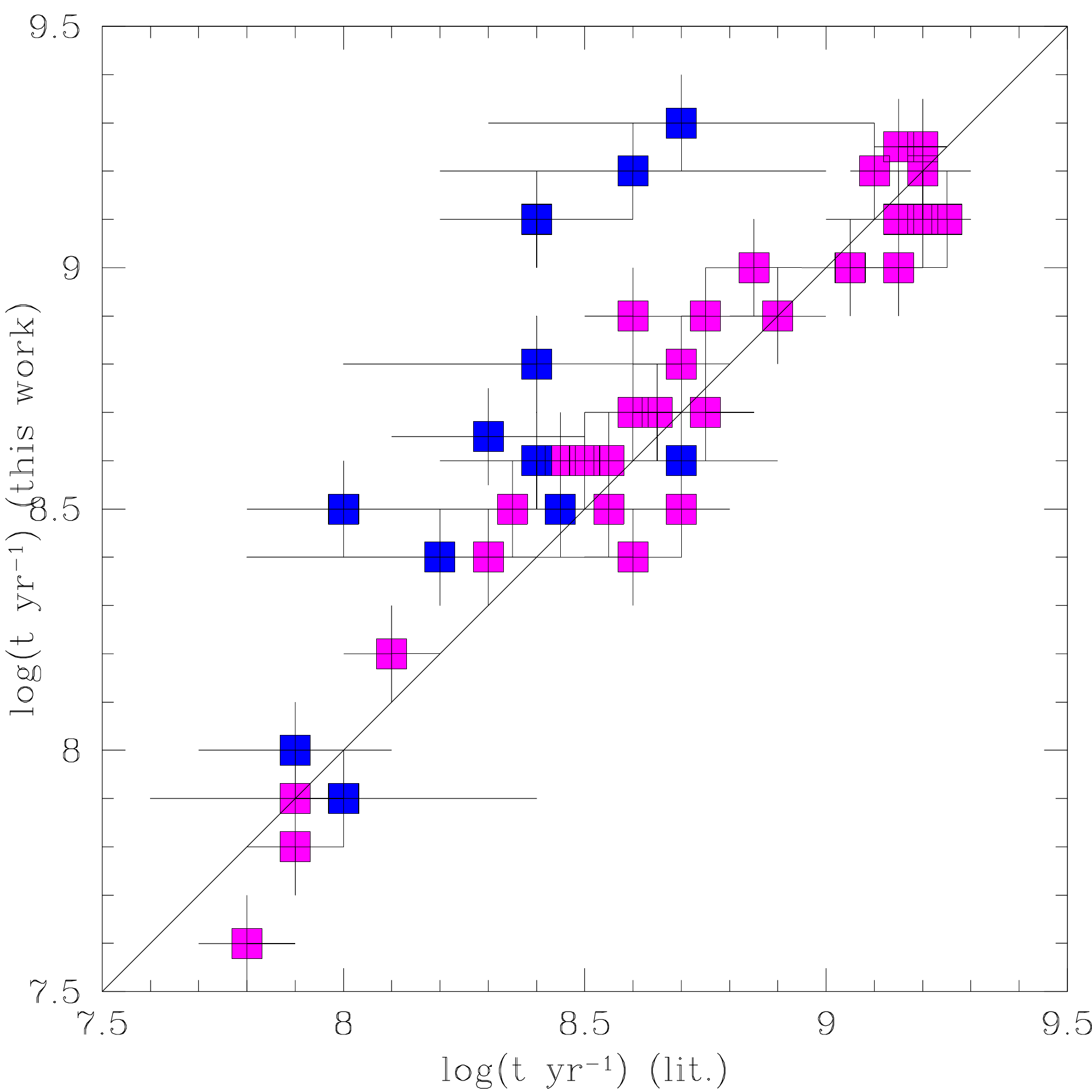}
    \caption{Comparison between ages taken from the literature and those
estimated in this work. Blue filled squares refer to clusters with
age estimated by \citet{getal10}.}
   \label{fig:fig3}
\end{figure}

We previously studied many of these clusters from different Washington photometry data sets,
while some others were analysed by \citet{getal10} as part of the 
Magellanic Cloud Photometric Surveys \citet[MCPS][]{zetal02}. The result of the comparison
between them is depicted in Fig.~\ref{fig:fig3}, while Table~\ref{tab:table3} lists
the values and references taken from the literature. As can be seen,  the agreement is fairly
good, except in the case of some clusters studied by Glatt et al. (blue filled squares). 
We recall that they did not perform any decontamination of field stars from the cluster CMDs
and that the MCPS reaches  MSTOs of clusters younger than $\sim$ 1 Gyr.
It is possible that the lack of cleaned CMDs and their shallower photometry
did not allow them to achieve more reliable age estimates. The very good agreement with
previous Washington photometry studies provides additional support to the nearly 30 per cent
of the clusters studied here for which, as far as we are aware,  their ages are estimated for 
the first time.

\begin{table}
\caption{Fundamental properties of the star cluster sample. }
\label{tab:table4}
\begin{tabular}{@{}llcccc}\hline
          & Cluster name &  $E(V-I)$ & \multicolumn{2}{c}{log($t$ yr$^{-1}$)} & Ref.\\
          &              &           & this work  & literature \\\hline
Field\,8  &   KMHK\,5    &  0.05         & 9.10$\pm$0.10 & 9.20$\pm$0.10  & 1 \\
Field\,9  &  NGC\,1644   &  0.05         & 9.10$\pm$0.10 & 9.15$\pm$0.15  & 1 \\
Field\,10 &  KMHK\,11    & 0.05          & 9.30$\pm$0.10 &   &  \\ 
Field\,11 & KMHK\,72     &  0.05         & 8.90$\pm$0.10  & 8.60$\pm$0.10  & 4 \\
Field\,12 & BSDL\,21  & 0.09 & 9.10$\pm$0.10  &     &  \\
          & BSDL\,38  & 0.06 &  8.60$\pm$0.10  & 8.70$\pm$0.20   & 2\\
          & BSDL\,75  & 0.09 &  8.90$\pm$0.10 &   &   \\
          & BSDL\,77  & 0.06 &  8.90$\pm$0.10  & 8.90$\pm$0.10   & 4\\ 
          & BSDL\,87  & 0.05 & 7.80$\pm$0.10   & 7.90$\pm$0.10  & 5\\
          & KMHK\,84  & 0.08 &  9.10$\pm$0.10 & 9.15$\pm$0.05  & 1\\
          & KMHK\,95  & 0.07 &  8.60$\pm$0.10  &  8.55$\pm$0.10  & 4\\
          & KMHK\,112 & 0.06 &  9.20$\pm$0.10 & 9.10$\pm$0.05 & 1\\
          & KMHK\,148 & 0.05 &  9.10$\pm$0.10  & 8.60$\pm$0.10   & 3\\
          & KMHK\,158 & 0.06 & 7.30$\pm$0.10  & 7.80$\pm$0.10 &5\\
          & new cluster& 0.06&   7.30$\pm$0.10 &  & \\
Field\,13 & BRHT\,45a &  0.09& 8.20$\pm$0.10& 8.10$\pm$0.10  & 4\\
          & BRHT\,45b &  0.09& 7.90$\pm$0.10  & 7.90$\pm$0.10 & 5 \\
          & BRHT\,62a &  0.05&  8.40$\pm$0.10 & 8.60$\pm$0.10 & 5\\
          & BRHT\,62b& 0.05 & 8.20$\pm$0.10  &  & \\
          & BSDL\,341 & 0.07 &  8.60$\pm$0.10 & 8.45$\pm$0.10  & 4\\
          & H88\,25  &0.07  &9.10$\pm$0.10   & 9.20$\pm$0.05 & 1\\
          & H88\,26   & 0.06 &8.90$\pm$0.10   & 8.90$\pm$0.10  & 3\\
          & H88\,32   &0.07  &8.60$\pm$0.10   &  8.40$\pm$0.10 &  5\\
          & H88\,40   & 0.08 & 9.00$\pm$0.10  &  8.85$\pm$0.10 & 3\\
          & H88\,52   &0.09  &  9.00$\pm$0.10 &  9.05$\pm$0.10 & 4\\ 
          & H88\,53   &0.09  &8.40$\pm$0.10   &  8.20$\pm$0.40  & 2\\
          & H88\,66   &0.07  & 8.80$\pm$0.10  &   & \\
          & H88\,67   & 0.07 & 9.00$\pm$0.10  &  9.15$\pm$0.10 & 1 \\
          & H88\,78   & 0.09 & 8.80$\pm$0.10  &   & \\
          & H88\,79   &0.08  &9.25$\pm$0.10   &  9.20$\pm$0.05 &1\\
          & KMHK\,286 & 0.09 &8.50$\pm$0.10   & 8.35$\pm$0.20 & 4\\
          & KMHK\,309s & 0.09 &   7.80$\pm$0.10& 7.90$\pm$0.10  & 5\\
          & KMHK\,333 &  0.09& 8.80$\pm$0.10  &  8.40$\pm$0.40  & 2\\
          & KMHK\,348&  0.08&  9.00$\pm$0.10 & 9.15$\pm$0.05 & 1\\
          & KMHK\,367& 0.08 &8.80$\pm$0.10   &  8.70$\pm$0.10 &3\\
          & KMHK\,390 &  0.07&  8.80$\pm$0.10 & 8.70$\pm$0.10 &3\\
          & NGC\,1764 &0.08  & 8.00$\pm$0.10  &7.90$\pm$0.10 & 5\\
          & NGC\,1786 & 0.06 &  10.10$\pm$0.10 & 10.10  & 6,7 \\
Field\,14 & BSDL\,527   &0.04 & 9.25$\pm$0.10 & 9.15$\pm$0.05  & 1\\
          & BSDL\,716   & 0.06& 8.70$\pm$0.10 &  8.60$\pm$0.10 &3\\
          & BRHT\,29a   &0.07 & 8.60$\pm$0.10 & 8.50$\pm$0.10  &4\\
          & BRHT\,29b   &0.07 & 7.80$\pm$0.10 & 7.90$\pm$0.10 &4\\
          & [HS66]\,131 & 0.09& 9.00$\pm$0.10 & 9.15$\pm$0.10  &4\\
          & KMHK\,505   &0.06 & 8.70$\pm$0.10 & 8.75$\pm$0.10  &4\\
          & KMHK\,506   &0.04 & 8.90$\pm$0.10 & 8.75$\pm$0.05 &3\\
          & KMHK\,531   &0.05 & 8.40$\pm$0.10 &8.30$\pm$0.10 &5\\
          & KMHK\,533   &0.04 & 9.20$\pm$0.10 & 9.10$\pm$0.05  &1\\
          & KMHK\,536   &0.06 & 8.50$\pm$0.10 & 8.55$\pm$0.10  &4\\
          & KMHK\,549   & 0.06& 8.70$\pm$0.10 &   &  \\
          & KMHK\,554   & 0.06& 8.00$\pm$0.10 &    &   \\
          & KMHK\,560   &0.04 & 9.00$\pm$0.10 & 9.15$\pm$0.05 & 1\\
          & KMHK\,586   &0.05 & 9.10$\pm$0.10 & 9.25$\pm$0.05 &1\\
          & NGC\,1829   &0.04 & 7.90$\pm$0.10  &   &  \\
          & NGC\,1838   &0.06 & 8.50$\pm$0.10 &  8.00$\pm$0.20    &2\\
          & ZHT\,AN\,8  &0.07 & 9.10$\pm$0.10 & &   \\
Field\,15 & BRHT\,34a   &0.11   &8.65$\pm$0.10 & 8.30$\pm$0.20  &2\\
          & BRHT\,34b  &0.11   &       8.50$\pm$0.10&  8.45$\pm$0.20    &2\\
          & BSDL\,1035  & 0.10  & 8.50$\pm$0.10  & 8.70$\pm$0.10 &3\\
          & BSDL\,1046  & 0.08  &9.20$\pm$0.10   & 8.60$\pm$0.40 &2\\
          & BSDL\,1139   &0.10   &8.55$\pm$0.10  &   &  \\
          & BSDL\,1141  &  0.11 &  8.30$\pm$0.10    &  \\
          & BSDL\,1152 & 0.10  &9.10$\pm$0.10   &  8.40$\pm$0.20   &2\\
          & BSDL\,1161 & 0.10&    9.00$\pm$0.10  &   &  \\
          & [HS66]\,221 & 0.08  & 8.40$\pm$0.10   & 8.20$\pm$0.20  &2\\\hline
\end{tabular}
\end{table}

\setcounter{table}{3}
\begin{table}
\caption{continued.}
\label{tab:table4}
\begin{tabular}{@{}llcccc}\hline
          & Cluster name &  $E(V-I)$ & \multicolumn{2}{c}{log($t$ yr$^{-1}$)} & Ref.\\
          &              &           & this work  & literature \\\hline
Field\,15 & ESO\,56-SC\,91&   0.10&  8.50$\pm$0.10&   &  \\
          & KMHK\,685   &0.11   & 9.30$\pm$0.10&  8.70$\pm$0.40 &     2\\
          & KMHK\,691   & 0.13  & 8.60$\pm$0.10&  8.40$\pm$0.20     & 2\\
          & KMHK\,727   &0.10   &9.20$\pm$0.10 & 9.20$\pm$0.10  & 8  \\
          & KMHK\,753   & 0.08  & 8.80$\pm$0.10   &  &  \\
          & KMHK\,775   & 0.11  & 8.00$\pm$0.10& 7.90$\pm$0.20    & 2\\
          & KMHK\,784   &0.14   &8.80$\pm$0.10   &   &  \\
          & KMHK\,794  &  0.09 &  7.80$\pm$0.10 &    &  \\
          & [SL63]\,327 &   0.07&  8.40$\pm$0.10 &      &  \\
          & [SL63]\,332 & 0.08  & 7.90$\pm$0.10   & 8.00$\pm$0.40    &2\\
          & [SL63]\,351 &  0.13 & 8.70$\pm$0.10  &8.65$\pm$0.05  &3\\
          & [SL63]\,370& 0.10&    9.00$\pm$0.10  &   &  \\\hline
\end{tabular}
\noindent Ref.: (1) \citet{p11a}; (2) \citet{getal10}; (3) \citet{p12c}; 
(4) \citet{chetal15}; (5) \citet{p14}; (6) \citet{brocatoetal1996};
(7) \citet{carrettaetal2000}; (8) \citet{piattietal2009}.
\end{table}

\section{analysis}

In order to study the cluster formation history along the minor axis of the LMC,
we constructed star cluster frequencies (CFs), defined as the number of 
clusters per age unit as a function of age.  CFs have the advantage that
they do not depend on the age interval, so that number of clusters at different
ages can be compared. We built CFs for Fields\,12 to  16, which are those that
contain more than one cluster. Each CF was normalized to the respective
total number of clusters for comparison purposes. From these CFs we 
compared the cluster formation activity in different epochs of the galaxy lifetime
for a particular field, and between different LMC fields as well.

\begin{figure*}
	\includegraphics[width=\textwidth]{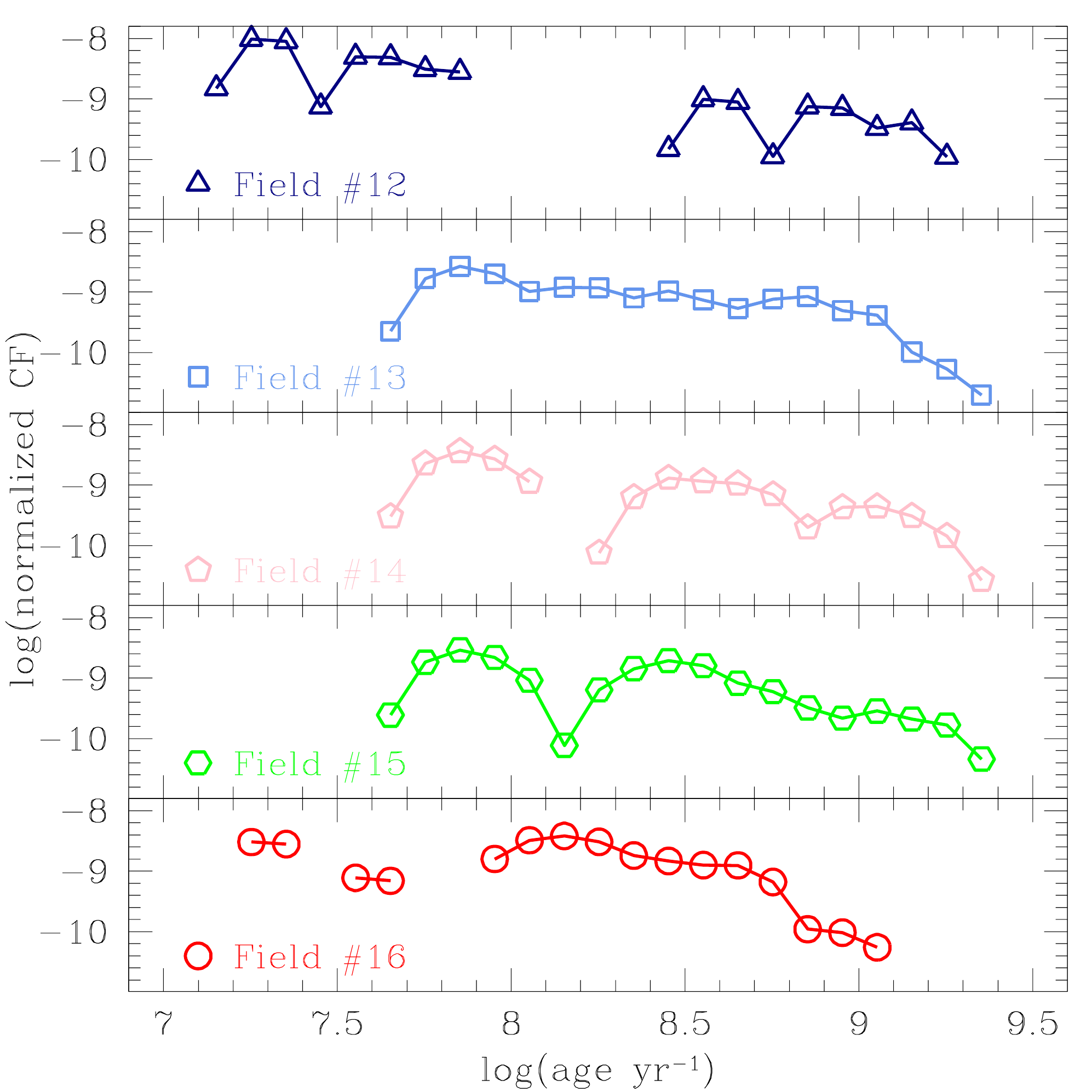}
    \caption{Normalized CFs of the observed LMC fields (see Fig.~\ref{fig:fig1})  plotted
    with different symbols as indicated in the bottom-left margin. The age intervals without clusters
    are distinguished as disconnected pair of points. Every CF has been divided (normalized) by the total
    number of clusters of the corresponding field, for comparison purposes.}
   \label{fig:fig4}
\end{figure*}

CFs were built assigning to each cluster age a
{\it Gaussian} distribution centred on the mean cluster age and with FWHM twice as big as 
the age uncertainty. Thus we dodge constructing age histograms that depend on the 
bin size and the end points of bins. Moreover,  since the age uncertainties can be  larger than 
 the size of the age  bins, a cluster  can actually reside in one of a few 
adjacent age  bins. We considered this effect while building an intrinsic CF.
We added all the {\it Gaussian} distributions and then computed the fraction of them that
fall in different age intervals ($\Delta$(log($t$ yr$^{-1}$)) = 0.1).
The result of summing the contribution of all {\it Gaussian} distributions
for each observed field is depicted in Fig.~\ref{fig:fig4}.  

At first glance, the oldest clusters turned out to be as old as log($t$ yr$^{-1}$) $\sim$ 9.3, 
in good agreement with the age range
of most of the LMC clusters \citep[log($t$ yr$^{-1}$) $\la$ 9.40, ][see their figure 6]{pg13},
with the exception of  ESO\,121-SC-03 (log($t$ yr$^{-1}$) $\sim$ 9.92) and 15 old 
globular clusters (log($t$ yr$^{-1}$) $\sim$ 10.1).  From log($t$ yr$^{-1}$) $\sim$ 9.3 until 
$\sim$ 8.5 each region has undergone a 
relatively similar increasing, nearly smooth, cluster formation activity, with two 
exceptions: Field\,12 (located near the edge of the LMC main body), which shows a 
slightly more interrupted cluster formation, and Field\,16  (located nearly at the 
LMC centre), for which clusters have later started to be formed (log($t$ yr$^{-1}$) $\sim$ 8.9).
Both pictures tell us about a particular spatial-age dependent scenario, in which clusters
started to be formed mostly throughout the LMC main body, except in the very innermost bar,
and that the gas out of which they formed was exhausted outside-in. The outside-in
formation scenario has been proposed previously by \citet{gallartetal2008,piattietal2009,meschin14}, 
among others.

The uncertain process of cluster disruption in the LMC tidal field may complicate the simplest interpretation
of the spatial patterns of cluster frequency versus age. As noted by \citet{lamers05}, the characteristic cluster
destruction timescale t$_d$ is a strong function of the ambient field density, with t$_d \propto \rho^{-1/2}$
in a typical model of the tidal disruption process. Since the field density varies dramatically across 
Fields 16 to 12, one might expect the age distribution to be steeper in the inner fields. However, all five
fields have roughly similar CF vs.\ age slopes (to within significant scatter). It is suggestive that Field 16
shows a steep decline of the CF at younger ages than the other four fields, but not conclusive based on these
data. In this context, it is worth noting that the cluster destruction timescale in the LMC as a whole has been
constrained to be longer than $\approx$1~Gyr \citep[e.g.,][]{pdg08}.

The process of cluster
formation apparently entered a quiescent stage during the period
log($t$ yr$^{-1}$) $\sim$ 8.0 -- 8.3 for most of the studied fields, while in the innermost bar region
(Field\,16) it reached its highest formation activity, ended at log($t$ yr$^{-1}$) $\sim$ 8.0.
Soon after (log($t$ yr$^{-1}$) $\sim$ 7.8--7.9), the regions where cluster formation had ceased or
gone to a quiescent stage, experienced a sudden episode of cluster formation activity that
notably continues until log($t$ yr$^{-1}$) $\sim$ 7.3 in the outermost one (Field\,12), 
although with interrupted short periods, and has been mirrored in the innermost one, Field\,16.

\citet{beslaetal07} and \citet{kallivayaliletal13} suggested that the first infall of the LMC to the Milky Way
(MW) took place at log($t$ yr$^{-1}$) $\sim$ 7.8, just close to the observed bursting
cluster formation of Fig.~\ref{fig:fig4}. If such a burst
were associated  with this close passage of the LMC, then we should expect that some amount of gas
to form young clusters reached the  outermost western regions of the galaxy. Note that only Field\,12
shows a roughly continuous cluster formation activity since log($t$ yr$^{-1}$) $\sim$ 7.9.
Recently, \citet{salemetal2015} and \citet{is2015} have suggested from observations and modelling of
H\,I spatial and velocity distributions, that the outer regions of the LMC were disturbed by
ram pressure effects due to the motion of the LMC in the MW halo. Particularly, \citet{salemetal2015}
found evidence that the LMC's gaseous disc has recently experienced ram pressure stripping, with
a truncated gas profile along the windward leading edge of the LMC disc. This means that some amount
of gas could travel in the  opposite direction to the LMC motion, and thus could contribute with
material for the cluster formation in the western-outermost part of the galaxy (see their figure\,17).  
Similarly, \citet{is2015} suggested possible outflows from the western LMC disc which could be
due to ram pressure. From these results, we speculate here with the possibility that the recent
star formation activity observed in the studied field could have triggered by tidal interaction
of the LMC during its first passage close to the MW.

\section{Conclusions}

In this work we analysed Washington $CMT_1$ photometry of star clusters located 
along the minor axis of the LMC. We covered a wide baseline in deprojected distances, 
from the very optical LMC centre up to $\sim$ 39 degrees outwards to the North-West.

We first performed an homogeneous search  for star clusters in the 17 36$\arcmin$$\times$36$\arcmin$ 
studied fields  using {\it Gaussian} and
{\it tophat} KDEs with a bandwidth of 0.4 arcmin to produce continuous stellar
density distributions, from which we identified stellar overdensities. 
For each of the  detected cluster candidates we built
CMDs statistically cleaned from field star contamination. The employed cleaning technique
makes use of variable cells that allows us to reproduce the field star CMD
as closely as possible. As a result, we confirmed the physical reality of 146 star
clusters located in the LMC main body -- two of them identified for the first time --, 
and concluded that an overall $\sim$
30 per cent of catalogued clusters in the surveyed regions are non-possible physical systems. 
We did not find any new cluster candidate in the outskirts of the LMC (deprojected  distance
$\ga$ 8  degrees), in very good agreement with a recent search for new clusters performed
on the SMASH\footnote{http://datalab.noao.edu/smash/smash.php} survey data base across
the Magellanic System \cite{p17}.

The confirmed clusters comprise a
complete sample, since we were able to detect any star cluster with stars from its
brightest limit down to its MSTO.
From matching theoretical isochrones to the cleaned cluster CMDs we estimated ages
taking into account the LMC mean distance modulus, the present day metallicity and the individual star
cluster colour excesses. As far as we are aware, these are the first age
estimates based on resolved stellar photometry for nearly 30 per cent of the
cluster sample. For the remaining clusters we found an excellent agreement between
ages estimated previously, most of them from Washington $CT_1$ photometry, and our derived
values. The derived ages are in the age range 7.2
< log($t$ yr$^{-1}$) $\le$ 9.4, in addition to the old globular cluster NGC\,1786. 

Finally, we  constructed CFs for each region with the aim of studying the cluster
formation history along the minor axis of the LMC. We confirmed that there exists
a  age-spatial dependence of the cluster formation activity, in the sense that clusters
in the innermost bar region started to be formed later (log($t$ yr$^{-1}$) $\sim$ 8.9) 
than those in  more outlying regions (log($t$ yr$^{-1}$) $\sim$ 9.3), which reinforces
previously proposed outside-in formation scenarios. Furthermore, when the cluster
formation apparently  entered a quiescent regime in most of the studied regions
(log($t$ yr$^{-1}$) $\sim$ 8.0 --8.3), the innermost bar experienced its highest
cluster formation activity. Later on, the outer studied fields show a sudden episode
of cluster formation (log($t$ yr$^{-1}$) $\sim$ 7.8-7.9), that continued until
log($t$ yr$^{-1}$) $\sim$ 7.3 only in the outermost LMC region, and had its mirrored
event in the innermost bar region. These outcomes could be the first evidence from
the study of star clusters that the first  passage of the LMC 
 by the Milky Way has triggered cluster formation  due to the ram pressure of MW halo gas.

\section*{Acknowledgements}
 We thank the referee for her thorough reading of the manuscript and
timely suggestions to improve it. 




\bibliographystyle{mnras}

\input{paper.bbl}




\newpage

\appendix

\section{Supplementary material provided online. }
\begin{table*}
\tiny
\caption{Photometric catalogue of stars measured in the field of KMHK\,685. 
Only a portion is shown
here for guidance regarding its form and content. The whole content is available online.
Photometric catalogues for all clusters in Table 4 are also available online.}
\begin{tabular}{@{}cccccccccccccc}\hline
ID  & R.A. & Dec. & X & Y & $C$ & $\sigma C$ & $M$ & $\sigma M$ & $T_1$ & $\sigma T_1$ & $\chi$ & sharpness & membership \\    
    & (hh:mm:ss)  & ($\degr$:$\arcmin$:$\arcsec$) & (pixels) & (pixels) & (mag) & (mag)&(mag) &(mag) &(mag) & (mag)& & & \\\hline 
- &- &- &- &- &- &- &- &- & -& -& -&- & - \\    
145713 & 05:14:46.694 &-68:20:50.40 &6090.440 & 973.777 & 20.4462  & 0.0290 & 20.1651 &  0.0410 & 19.9675 &  0.0420 &  0.7520  & 0.2670  & 2\\
145718 & 05:14:50.961 &-68:20:50.26 &6090.743 &1063.417 & 20.3552  & 0.0160 & 19.4249 &  0.0190 & 18.8257 &  0.0150 &  1.1603  & 0.3897  & 2\\
145730 & 05:14:56.604 &-68:20:50.04 &6091.319 &1181.795 &20.3062  & 0.0320 & 19.3921 &  0.0120 & 18.8368 &  0.0200 &  0.9477  & 0.0187  & 1\\
- &- &- &- &- &- &- &- &- & -& -& -&- & - \\\hline

 \end{tabular}
\end{table*}




\bsp	
\label{lastpage}
\end{document}